\documentclass[a4paper,twoside]{article}

\usepackage{epsfig}
\usepackage{subcaption}
\usepackage{calc}
\usepackage{amssymb}
\usepackage{amstext}
\usepackage{amsmath}
\usepackage{amsthm}

\usepackage{multicol}
\usepackage{pslatex}
\usepackage{apalike}
\usepackage{algorithm2e}
\usepackage[bottom]{footmisc}
\usepackage{url}
\usepackage{blindtext}
\usepackage{hyperref}
\usepackage{nicefrac}
\usepackage{braket}
\usepackage{mathtools}
\usepackage{bbold}
\usepackage[T1]{fontenc}
\usepackage{orcidlink}
\usepackage{float}

\DeclareMathOperator*{\argmin}{arg\,min}

\usepackage{SCITEPRESS}     

\begin{document}

\title{Improving Parameter Training for VQEs by Sequential Hamiltonian Assembly}

\author{\authorname{Anonymous Author(s)}}

\author{\authorname{Jonas Stein\sup{1}\orcidlink{0000-0001-5727-9151}, Navid Roshani\sup{1}, Maximilian Zorn\sup{1}\orcidlink{0009-0006-2750-7495}, Philipp Altmann \orcidlink{0000-0003-1134-176X}, \\Michael Kölle\sup{1}\orcidlink{0000-0002-8472-9944} and Claudia Linnhoff-Popien\sup{1}\orcidlink{0000-0001-6284-9286}} 
\affiliation{\sup{1}LMU Munich, Germany}
\email{jonas.stein@ifi.lmu.de}
} 


\abstract{A central challenge in quantum machine learning is the design and training of parameterized quantum circuits (PQCs). Similar to deep learning, vanishing gradients pose immense problems in the trainability of PQCs, which have been shown to arise from a multitude of sources. One such cause are non-local loss functions, that demand the measurement of a large subset of involved qubits. To facilitate the parameter training for quantum applications using global loss functions, we propose a Sequential Hamiltonian Assembly, which iteratively approximates the loss function using local components. Aiming for a prove of principle, we evaluate our approach using Graph Coloring problem with a Varational Quantum Eigensolver (VQE). Simulation results show, that our approach outperforms conventional parameter training by 29.99\% and the empirical state of the art, Layerwise Learning, by 5.12\% in the mean accuracy. This paves the way towards locality-aware learning techniques, allowing to evade vanishing gradients for a large class of practically relevant problems.} 

\onecolumn \maketitle \normalsize \setcounter{footnote}{0} \vfill

\section{\uppercase{Introduction}}
\label{sec:introduction}
One of the most promising approaches towards an early quantum advantage is quantum machine learning based on parameterized quantum circuits (PQCs)~\cite{Cerezo2021}. PQCs are generally regarded as the quantum analog to artificial neural networks, as they resemble arbitrary function approximators with trainable parameters~\cite{PhysRevA.103.032430}. Mathematically, PQCs are parameterized linear functions that live in an exponentially high dimensional Hilbert space with respect to the number of qubits involved. In essence, the ability to efficiently execute specific computations in this large space allows for provable quantum advantage~\cite{10.1145/237814.237866,deutsch1992rapid}.

A core difference in a gradient based training process of classical ANNs with PQCs, is the efficiency: While the gradient calculation is invariant in the number of parameters in classical ANNs, its runtime complexity evidently scales linearly with the number of parameters for PQCs~\cite{PhysRevA.98.032309}. As a gradient is merely the expectation value of the probabilistic measurement from a quantum circuit, it's error-dependent runtime scaling is $\mathcal{O}\left(1/\epsilon\right)$ compared to the classical $\mathcal{O}\left(\log\left(1/\epsilon\right)\right)$~\cite{knill2007optimal}. This disadvantage manifests substantially in case of vanishing gradients, which are a common problem in PQCs~\cite{McClean2018}, especially as the gradients can vanish exponentially in the number of qubits~\cite{McClean2018}, as opposed to the number of layers in the classical case~\cite{bradley2010learning,pmlr-v9-glorot10a}.

Subsequently, much attention was devoted to exploring causes of vanishing gradients for PQCs by the scientific community, which identified the four following possible causes of vanishing gradients:
\begin{enumerate}
    \item \textbf{Expressiveness}: The larger the reachable subspace of the Hilbert space, the more likely gradients can vanish \cite{PRXQuantum.3.010313}.
    \item The \textbf{locality} of the measurement operator associated with the loss function: The more qubits have to be measured, the more likely gradients can vanish \cite{cerezo2021cost,uvarov2021barren,kashif2023impact}.
    \item The \textbf{entanglement} in the input: The more entangled or random the initial state, the more likely gradients can vanish \cite{McClean2018,cerezo2021cost}.
    \item Hardware \textbf{noise}: The more noise and the more different noise types present in hardware, the more likely measured gradients can vanish \cite{wang2021noise,stilck2021limitations}.
\end{enumerate}
More recently, mathematical approaches unifying the theory underlying causes were proposed, allowing for a quantification of the presence of vanishing gradients in a given PQC~\cite{ragone2023unified,fontana2023adjoint}. Similar to the vanishing gradients problem in classical machine learning, many techniques are being investigated in related work (see~\cite{ragone2023unified} for an overview). Some concrete examples to this are clever parameter initialization~\cite{NEURIPS2022_7611a3cb} and adaptive, problem informed learning rates~\cite{PRXQuantum.3.020365}.

In this article we propose another approach to facilitate efficient parameter training in PQCs coined \emph{Simulated Hamiltonian Assembly} (SHA), which is targeted towards the issue of locality. In essence, we propose to exploit the structure of most practically employed measurement operators $\hat{H}$, i.e., $\hat{H}=\sum_i \hat{H}_i$ where $\hat{H}_i$ is a \emph{local} Hamiltonian for all $i$. In this context, locality means that the given operator only acts non-trivially on a small subset of all qubits, such that the loss is merely influenced by the outcome of a small subset of qubits. An important class of problems exemplifying the property of having a local decomposition while also being a promising contender towards quantum advantage are combinatorial optimization problems~\cite{10.3389/fphy.2014.00005,PhysRevX.8.031016,pirnay2023inprinciple}.

Inspired from iterative learning approaches like layerwise learning from (quantum) machine learning and iterative rounding from optimization, SHA starts with a partial sum of the measurement operator $\hat{H}$ (e.g., simply $\hat{H}_1$) and iteratively adds more terms until it completely \emph{assembled} the original measurement operator $\hat{H}$. This iterative approximation of the loss function allows to start with a local measurement operator, which increases the ease of finding good initial parameters outside potential barren plateaus (i.e., areas in which the gradient vanishes), and subsequently aims to continually evade barren plateaus until the complete, typically global, measurement operator is used.

As a proof of principle, we conduct a case study for the problem of graph coloring using the state of the art PQCs based approaches to do so: The Variatonal Quantum Eigensolver (VQE) and the Quantum Approximate Optimization Algorithm (QAOA). The problem of graph coloring is chosen, as it has a complex loss function which challenges standard parameter training approaches and furthermore allows for a comparison of different assembly approaches. Our evaluation shows a significant improvement in solution quality when using SHA compared to standard gradient descent based training, as well as comparable state of the art approaches from related work.

The remainder of this paper is structured as follows. Section~\ref{sec:background} provides necessary theory for training parameters in PQCs, as well as the intuitions underlying the VQE and QAOA. Subsequently, section~\ref{sec:relatedwork} discusses two different approaches from related work, that can be used to accelerate parameter training. In section~\ref{sec:sha}, we present the theory behind our main contribution: Sequential Hamiltonian Assembly (SHA). The experimental setup used for evaluation is then described in section~\ref{sec:experiments}, before evaluating the results in section~\ref{sec:evaluation}. Finally, we conclude our findings and propose possible future work in section~\ref{sec:conclusion}.

\section{\uppercase{Background}}
\label{sec:background}
In this section, we present the fundamental theory behind training parameters in PQCs, and also introduce the algorithms used for evaluation.

\subsection{Training parameterized quantum circuits}
\label{subsec:prametertraining}
Similar to classical machine learning, most practically employed parameter training techniques for PQCs rely on gradient based methods. The cornerstone for calculating the gradients of a PQCs $U\left(\theta,x\right)$, where $U$ is a unitary matrix acting on all $n$ qubits, $x\in\mathbb{C}^k$ denotes the data input and $\theta\in\mathbb{R}^m$ the parameters, is the \emph{Parameter Shift Rule}. Exploiting that all PQCs can be decomposed into possibly parameterized single qubit gates and non-parameterized two qubit gates~\cite{nielsen_chuang_2010}, the parameter shift rule takes advantage of the periodic nature of single qubit gates. Similar to how $\nicefrac{d}{dx} \sin\left(x\right) = \sin\left(x+\nicefrac{\pi}{2}\right)$, one can show that $\forall i$:
\begin{align} \label{eq:parametershift}
    \dfrac{\partial}{\partial \theta_i} U\left(\theta,x\right)\ket{\psi} = \dfrac{U\left(\theta^+,x\right)\ket{\psi} - U\left(\theta^-,x\right)\ket{\psi}}{2},
\end{align}
where $\ket{\psi}$ denotes an arbitrary initial state and $\theta^\pm\coloneqq \left(\theta_1,...,\theta_{i-1},\theta_i\pm\nicefrac{\pi}{2},\theta_{i+1},...,\theta_m\right)$~\cite{PhysRevA.98.032309,PhysRevA.99.032331}. This allows utilizing the original PQC $U$ to calculate the gradient as efficiently as the forward pass for each parameter, which can be parallelized using multiple QPUs to achieve the same runtime complexity as the backward pass in classical ANNs, when neglecting error. 

\subsection{The Variational Quantum Eigensolver}
\label{subsec:VQE}
The Variational Quantum Eigensolver is a quantum optimization algorithm that utilizes a PQC $U\left(\theta\right)$ to approximate the \emph{ground state} of a given Hamiltonian $\hat{H}$, i.e., an eigenvector of $\hat{H}$ corresponding to its smallest eigenvalue~\cite{Peruzzo2014}. The VQE stems from the \emph{variational method}, which describes the process of iteratively making small changes to a function (in our case $f:\theta \mapsto U\left(\theta\right)\ket{0}$) to approximate the $\argmin$ of a given function (in our case $g:\ket{\varphi} \mapsto \bra{\varphi}\hat{H}\ket{\varphi}$)~\cite{lanczos2012variational}.

While the original proposition of the VQE is focused on solving chemical simulation problems, the algorithm can more generally be used to solve arbitrary combinatorial optimization problems by executing the following steps:
\begin{enumerate}
    \item Encode the domain space of the given combinatorial optimization function $h:X\rightarrow \mathbb{R}$ in binary, i.e., define a map $e:X\rightarrow \left\{0,1\right\}^n$. This allows for the equality $h(x)= \bra{e(x)}\hat{H}\ket{e(x)}$ by defining $\hat{H}$ as a diagonal matrix with eigenvalues $h(x)$ corresponding to the eigenvectors $\ket{e(x)}$, such that finding the the ground state of $\hat{H}$ corresponds to finding the global minimum of $h$.
    \item Select a quantum circuit architecture defining the function approximator $U\left(\theta\right)$.
    \item Choose an initial state $\ket{\psi}$ (typically $\ket{0}$ to avoid additional state preparation) as well as initial parameters $\theta$ (e.g., $\theta_i=0$ $\forall i$).
    \item Specify an optimizer for parameter training.
\end{enumerate}
In practice, even though much scientific exploration has already been conducted~\cite{Du2022,sim2019expressibility}, the selection of a suitable and efficient circuit architecture appears to be most challenging step.

\subsection{The Quantum Approximate Optimization Algorithm}
The Quantum Approximate Optimization Algorithm can be regarded as a special instance of the VQE, that utilizes the principles of Adiabatic Quantum Computing (AQC) to construct provably productive circuit architectures~\cite{farhi2014quantum}. AQC is paradigm to conduct quantum computing that is equivalent to the standard quantum gate model~\cite{10.1109/FOCS.2004.8} and is motivated by the \emph{adiabatic theorem}. The adiabatic theorem essentially states, that a physical system remains in its eigenstate  when the applied time evolution is carried out slowly enough~\cite{Born1928}. Exploiting the equivalence of solving a ground state problem and combinatorial optimization described in section~\ref{subsec:VQE}, AQC can be used to solve combinatorial optimization problems.

Mathematically, a computation in AQC can be described by a time dependent Hamiltonian $\hat{H}(t)=\left(1-t\right)\hat{H}_M + t\hat{H}_C$, where time $t$ evolves from $0$ to $1$, $\hat{H}_M$ denotes the Hamiltonian whose ground state corresponds to the initial state, and $\hat{H}_C$ is defined such that it corresponds to the given optimization problem. As it is straightforward to prepare an initial state corresponding to the ground state of a Hamiltonian (e.g., $\ket{+}^{\otimes n}$ wrt. $\hat{H}_M\coloneqq - \sum_{i=1}^{n} \sigma^x_i$), and as the Hamiltonians $\hat{H}_C$ corresponding to practically relevant combinatorial optimization problems can be decomposed into a sum of at most polynomially many local Hamiltonians, AQC can be utilized to efficiently solve many optimization problems.

In essence, the QAOA simulates the continuous time evolution of AQC described above by using discretization techniques, as computations in the standard model of quantum computing are conducted using purely discrete quantum gates. As implied by the adiabatic theorem, the maximally allowed time evolution speed merely depends on the difference between the smallest and the second-smallest eigenvalue of $\hat{H}(t)$ at each point in time $t$. Aiming to exploit this possibility of accelerating the time evolution beyond the maximally possible constant speed, the QAOA introduces parameters that can essentially vary the speed of time evolution. A careful mathematical derivation of these ideas yields the following PQC:
\begin{equation*}
    U\left(\beta, \gamma\right) = U_M(\beta_p) \cdot U_C(\gamma_p) \cdot \ldots \cdot U_M(\beta_1) \cdot U_C(\gamma_1) \cdot H^{\otimes n}
\end{equation*}
where $U_M(\beta_i) \coloneqq e^{-i\beta_i \widehat{H}_{M}}$, $U_C(\gamma_i) \coloneqq e^{-i\gamma_i \widehat{H}_{C}}$, and $p\in\mathbb{N}$, such that $U\left(\beta, \gamma\right)$ approaches AQC for $p\rightarrow \infty$, and constant speed, i.e,  $\beta_i = 1-\nicefrac{i}{p}$, and $\gamma_i = \nicefrac{i}{p}$.

In practice, the QAOA (incl. slight adaptations of it) often yields state of the art results compared to other quantum optimization methods~\cite{blekos2023review}. Nevertheless, its runtime complexity strongly depends on how many, and how local the Hamiltonians $\hat{H}_i$ composing $\hat{H}_C=\sum_i \hat{H}_i$ are, as well as how natively they fit on the given hardware topology. Due to the repetitive application of $U_C$, this restricts the number of usable discretization steps $p$ significantly, and therefore the solution quality, as it grows proportional to $p$. For this reason, other VQE-based PQCs can regularly outperform the QAOA on near term quantum computers, in spite of its property of provably finding the optimal solution when given enough time~\cite{9669165,Skolik2021}.

\section{\uppercase{Related Work}}
\label{sec:relatedwork}
To allow for a comparison of our approach with other methods to enhance parameter training in (VQE-based) PQCs, we now provide a short introduction into two prominent techniques: Layerwise learning and Layer-VQE. To the best knowledge of the authors, no other baselines have been proposed, that are more similar to our methodology in terms of iteratively guiding the parameter learning process while aming to evade barren plateaus.

\subsection{Layerwise Learning}
\label{subsec:ll}
Inspired by classical layerwise pretraining strategies used in deep learning (for reference, see \cite{NIPS2006_5da713a6}), \cite{Skolik2021} showed, that iteratively training a subset of the parameters in PQCs can significantly improve the solution quality. As an input to their approach, they assume a layered-structure of the PQC, which is common in most of the literature. The training is then performed in two phases. In the first phase, the parameters are trained while sequentially assembling the PQC in a layerwise manner:
\begin{enumerate}
    \item Start with a PQC consisting of the first $s$-layers of the given PQC, while initializing all parameters to zero, aiming to initialize close to a barren plateau\footnote{A barren plateau is a region in the parameter landscape, in which the gradients vanish.} free identity operation.
    \item Train the parameters for a predefined number of maximal optimization steps.
    \item Add the next $p$-layers to the PQC and fix the parameters of all but the last added $q$-layers, while initializing all new parameters to zero.
    \item Train the parameters of the last added $q$-layers for a predefined number of maximal optimization steps.
\end{enumerate}
This process is repeated until the addition of new layers does not improve the solution quality or a specified maximal depth is reached. Note, that all introduced variables (i.e., $s$, $p$ and $q$) are hyperparameters that potentially need to be trained in order to fit the specific requirements of the problem instances of interest.

In the second phase, another round of parameter training, now with the fully assembled circuit, is conducted. Here, a fixed fraction of layers $r$ is trained in a sliding window manner, while fixing the rest of the circuits parameters. Each contiguous subset of layers is trained for a fixed number of maximal optimization steps.

By choosing the number of optimization steps for each part of the training sufficiently low, overfitting can be obviated, while also bounding the overall training duration from above. In subsequent papers, it has been shown that there is a lower bound on the size of the subset of simultaneously trained layers in layerwise learning, to allow for effective training~\cite{PhysRevA.103.032607}. Nevertheless, there are relevant applications such as image classification, for which experimental results indicate significantly lower generalization errors when using layerwise learning~\cite{Skolik2021}.

\subsection{Layer-VQE}
Building upon the insights gained in \cite{Skolik2021} and \cite{PhysRevA.103.032607} (see section~\ref{subsec:ll}), \cite{9669165} proposed the iterative parameter training approach \emph{Layer-VQE}, that essentially resembles a special case of layerwise learning. The core idea for Layer-VQE is that each layer must equal an identity operation when setting the parameters to zero. This ensures that the output state of the circuit does not change when adding a new layer, so that the search in the solution space of the given optimization problem continues from the previously optimized solution. The other fundamental specification of layerwise learning in Layer-VQE is that the second phase is omitted (i.e., $r=0$), as $q$ is chosen to be cover all previously inserted layers. This means that instead of merely training the $q$ last layers, all inserted layers are trained simultaneously. To limit the number of new parameters in each step, only one layer is added in each iteration (i.e., $p=1$). Beyond these substantive specifications of layerwise learning, a small detail is added in Layer-VQE: An initial layer consisting of parameterized $R_y$ rotations on every qubit.

Based on the large scale evaluation conducted in \cite{9669165}, Layer-VQE can outperform QAOA in terms of solution quality and circuit depth for specific optimization problems, especially on noisy hardware.

\section{\uppercase{Sequential Hamiltonian Assembly}}
\label{sec:sha}
We now propose our core contribution in this article: The Sequential Hamiltonian Assembly (SHA) approach, targeted towards facilitating parameter training of PQCs on global cost functions. Inspired by the concept of iteratively guiding the learning process by sequentially assembling the quantum circuit of a VQE layer by layer, as done in layerwise learning, we propose to instead assemble the often global Hamiltonian $\hat{H}=\sum^N_{i=1}\hat{H}_i$ (i.e., the cost function), by iteratively concatenating its predominantly local components $\hat{H}_i$. A significant motivation of this approach are similar, very successful strategies proposed in combinatorial optimization, that start with a relaxed version of the cost function and iteratively remove imposed relaxations to reassemble the original cost function (for a review on one of these approaches called iterative rounding, see~\cite{bansal:LIPIcs:2014:4827}).

In the following, we provide steps concretizing the proposed concept of SHA for a given PQC and a decomposable Hamiltonian $\hat{H}=\sum^N_{i=1}\hat{H}_i$:
\begin{enumerate}
    \item Select a strategy on which Hamiltonians should be added in which iteration, i.e., a partition $P=\left\{P_1,...,P_M\right\}$ where $P_i\subseteq\left\{1,...,N\right\}$ such that $\bigcup_{j=1}^M P_j = \left\{1,...,N\right\}$.
    \item Choose a maximal number of parameter optimization steps per iteration $s_j$, optimally low enough to avoid overfitting.
    \item Iteratively optimize the parameters of the given PQC wrt. the Hamiltonian $\sum_{i\in \bigcup_{j=1}^k P_j} \hat{H}_i$ for each $k\in\left\{1,...,N\right\}$ in ascending order for max. $s_k$ steps.
\end{enumerate}

As shown in the evaluation section, the assembling strategy can have a significant impact in the solution quality. For properly evaluating this, we propose three different approaches:
\begin{itemize}
    \item \textbf{Random}: Use equally sized, non-overlapping partitions, while the partition assigned to each $\hat{H}_i$ is random.
    \item \textbf{Chronological}: Use equally sized, non-overlapping partitions, while assigning the Hamiltonians in the order specified upon input.
    \item \textbf{Problem inspired}: Use a problem inspired partitioning, where the terms in each partition share a common, problem specific, property.
\end{itemize}
Note that in practice, given Hamiltonians $\hat{H}$ can be divided into many sub-Hamiltonians $\sum^N_{i=1}\hat{H}_i$, such that it might take too long to progress with one $\hat{H}_i$ at a time. For our purposes, and computational restrictions, $M\leq 10$ already showed decent results. We choose these three approaches, as they provide different degrees of information on the problem instance at hand: While \emph{random} does not provide any information, \emph{chronological} does do so in many cases (see examples in~\cite{10.3389/fphy.2014.00005}). Finally, in the \emph{problem inspired} approach, all available knowledge of the problem instance can be used to solve it in an iterative manner, as shown in the following example using graph coloring.

Graph coloring is a satisfiability problem concerning the assignment of a color to each node, such that no neighboring nodes share the same color. While there also exists an optimization of it, i.e., finding the smallest number of colors, so that such a color assignment is still possible, we focus on the satisfiability version, as it already inherits complex structural properties and is easier to evaluate. To save computational resources, we employ the Hamiltonian formulation proposed in~\cite{9259934}, which needs the least amount of informationally required qubits to solve this problem:
\begin{equation*}
    \sum_{(v,w)\in E}\sum_{a \in \mathbb{B}^m}\prod_{l=1}^m\left(\mathbb{1}+(-1)^{a_l}\sigma_{v,l}^z\right)\left(\mathbb{1}+(-1)^{a_l}\sigma_{w,l}^z\right)
\end{equation*}
Note, that this formulation only works when considering problems where the number of colors $k$ equals a power of two (i.e., $\exists m\in\mathbb{N}:2^m=k$), which we do in our evaluation ensuring minimal requirements regarding computational resources.

This Hamiltonian can be decomposed into at most $\left|E\right|\cdot k$ different Pauli terms with at most $\left|E\right|\cdot \binom{m}{l}$ many $l$-local terms, where $l\in\left\{1,...,m\right\}$. To decompose this substantial number of local sub-Hamiltonians, we propose a \emph{nodewise} approach, i.e., $\left|V\right|$ many partitions $P_j$ that contain all Pauli term indices involving the node $v_j\in V$. The results for this approach subsequently show, that the increase in problem instance information can improve the solution quality significantly. Therefore SHA demonstrates an example of how the problem with training with respect to global cost functions can be approached by iteratively assembling them from local subproblems in a problem informed manner.

\section{\uppercase{Experimental Setup}}
\label{sec:experiments}
In this section, we motivate and describe our choice of problem instances, PQC architectures, and hyperparameters, which are used in the subsequently following evaluation.

\subsection{Generating Problem Instances}
To generate bias-free, statistically relevant, problem instances, we take the standard approach of using random graphs from the Erdős-Rényi-Gilbert model \cite{10.1214/aoms/1177706098}. This model conveniently allows for the generation of graphs with a fixed number of nodes while varying the number of edges, such that graphs of different hardness can be generated. Generally, the hardness of solving the graph coloring problem for a fixed number of colors in a random graph is proportional to the number of edges: The more edges, the harder~\cite{PhysRevE.76.031131}. To quantify the hardness, we straightforwardly use the percentage of correct solutions in the search space, as commonly done for satisfiability problems.

The dataset resulting from these considerations is displayed in table~\ref{tab:graphs}, where $p$ denotes the probability of arbitrary node pairs to be connected by an edge, $r$ the percentage of correct solutions in the search space, and $s$ is the absolute number of correct solutions in the search space. To achieve a sensible tradeoff between computational effort needed for simulation and reasonably sized graphs, all instances considered involve 8 nodes, and 4 colors, which amounts to $8 \cdot \log_2 \left(4\right)=16$ qubits and a reasonably large search space of size $4^8 = 65535$. As tested in course of the evaluation and in line with previously mentioned theoretic arguments, these graphs vary in their difficulty according to $p$ and $s$, with graph 9 being the hardest by far, and 5 being the easiest.

\begin{table}
    \centering
    \begin{tabular}{c||c|c|c|c}
        Graph id & $p$ & $r$ & $s$ & seed \\ \hline \hline
        1  & $0.30$ & $1.025\%$ & 672 & 7\\
        2  & $0.55$ & $1.501\%$ & 984 & 8\\
        3  & $0.40$ & $1.025\%$ & 672 & 9\\
        4  & $0.40$ & $1.428\%$ & 936 & 10\\
        5  & $0.35$ & $3.369\%$ & 2208 & 11\\
        6  & $0.30$ & $2.051\%$ & 1344 & 12\\
        7  & $0.35$ & $3.223\%$ & 2112 & 13\\
        8  & $0.50$ & $0.879\%$ & 576 & 14\\
        9  & $0.90$ & $0.037\%$ & 24 & 15\\
        10 & $0.40$ & $0.659\%$ & 432 & 16\\
    \end{tabular}
    \caption{Utilized graph problem instances, generated using the \texttt{fast\_gnp\_random\_graph} function from \texttt{networkx} \cite{hagberg2008exploring}. Every graph was checked to be fully connected.}
    \label{tab:graphs}
\end{table}

\subsection{Selecting suitable circuit layers}
Aiming to test our approach for a wide variety of different PQC architectures, we draw from the extensive list provided in~\cite{https://doi.org/10.1002/qute.201900070}, which contains many structurally differently PQCs.
To extend these architectures to the required number of 16 qubits, the underlying design principles are identified and extended to cover all 16 qubits (i.e., ladder, ring, or triangular entanglement layers, as well as single-qubit rotation layers). As testing all 19 proposed circuit architectures for such a large amount of qubits would exceed the computational simulation capacities available to the authors, we conducted a small prestudy to select a suitable subset according to the following criteria: (1) significantly better-than-random performance, (2) limited number of parameters (to reduce training time), and (3) variance in the architectures. Eliminating circuit 9 for reason (1), circuits 5 and 6 for reason (2) and dropping circuits (2 \& 4), (8, 12 \& 16), and (13 \& 18) for reason (3)\footnote{For the elimination of all but one circuit with respect to reason (3), the one with the best performance was selected.}, circuits 1, 3, 8, 12, 13, 16 and 18 are used in our evaluation. A small prestudy to the evaluation showed, that these circuits display roughly similar solution qualities when averaged over all problem instances, only with circuit 12 performing slightly worse, which might be induced by its layers' non-identity property when zeroing all parameters.

\subsection{Hyperparameters}
As outlined in sections~\ref{sec:background}, \ref{sec:relatedwork}, and~\ref{sec:sha}, the baselines, as well as SHA possess important hyperparameters, for which we now specify concrete values. As almost all approaches require layerwise structured PQCs, we conducted a small prestudy on the number of circuit layers needed for solving the given problem instances. As a result of this, we choose to conduct our case study using three layers, as more layers did not improve the solution quality significantly. For layerwise learning, we thus choose $s=1$, $p=1$, $q=1$, to train only a single layer at a time in the first phase and $r=1$ to train the full PQC in the second phase. Any higher values for $s$, $p$ and $q$ would hardly be sensible, as the number of layers is merely three. With $r=1$, we use the most potent approach according to~\cite{PhysRevA.103.032607}.

For all approaches, we employed the COBYLA optimizer~\cite{COBYLA_1}, due to its empirically proven highly efficient performance for similarly sized problems~\cite{Joshi_2021,9345015}. While setting the number of maximally possible optimization steps to 4000 for all runs, we limit the maximum learning progress for the training steps in which only a subset of the parameters where trained, by setting the least required progress in each optimization step to 0.8. For the final steps, in which all parameters are trained concurrently, this variable is set to $10^{-6}$, as overfitting is not a concern here anymore. Finally, a shot based circuit simulator was used to account for shot-based, imperfect estimation values present when using quantum hardware. For all circuit executions, the number of shots was set to 200, which already allowed for reasonably good results.

\section{\uppercase{Evaluation}}
\label{sec:evaluation}
To evaluate our approach, we first compare the proposed assembly strategies and then review the solution quality of SHA. Further, we show that SHA can be productively combined with the other discussed quantum learning methods layerwise learning and Layer-VQE. Finally, we compare the time complexity of all discussed approaches. To guarantee sufficient statistical relevance, all experiments are averaged over five seeds.

\subsection{Comparing Assembly Strategies}
\begin{figure}[ht]
    \centering
    \includegraphics[width=\columnwidth]{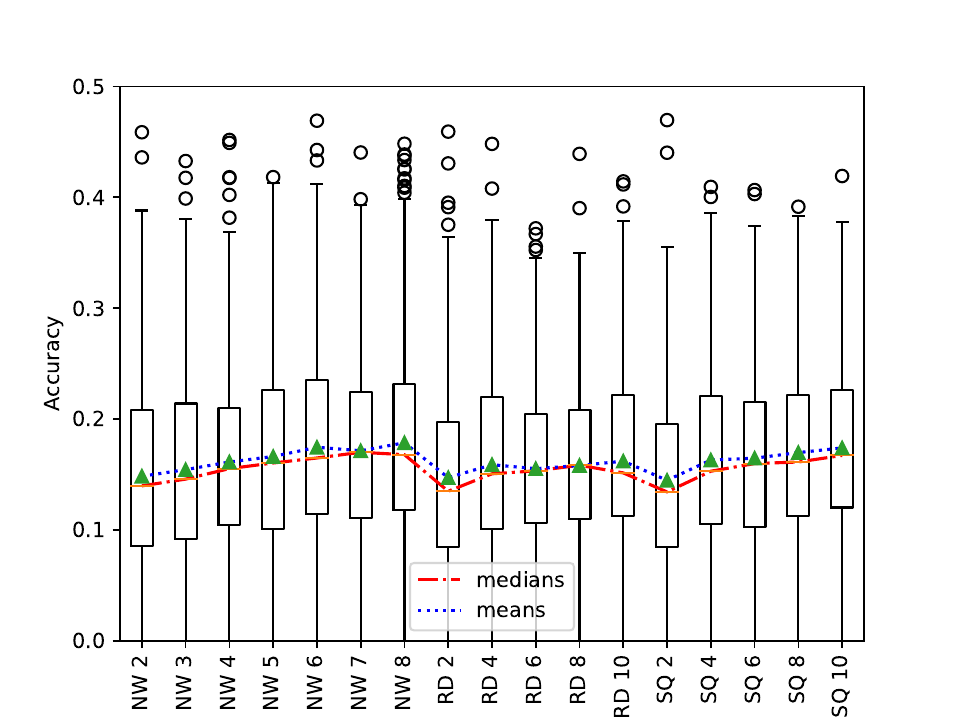}
    \caption{Accuracy over all graphs and circuit architectures per used assembly strategy.}
    \label{fig:assembly-strategies}
\end{figure}

As described in section~\ref{sec:sha}, SHA depends on the chosen Hamiltonian assembly strategy. Depending on the information available on the specific problem instance and Hamiltonian sum, we now evaluate the three different strategies proposed: (1) random, (2) chronological, and (3) problem inspired. Examining the results displayed in figure~\ref{fig:assembly-strategies}, we can see that random (RD $i$) performed the worst, when partitioning into $i\in\{2,4,6,10\}$ many equally sized partitions. The chronological approach (SQ $i$) performed reasonably better and even came close to the performance of the problem informed, nodewise (NW $j$) strategy, where $j\in\{2,...,8\}$ denotes the number of connected subgraphs used in the assembly. From the presented results, we can clearly observe, that a problem instance informed strategy leads to better results. Based on the results of the chronological approach, we can also see an explicit progression in the sense that, the more pronounced the structure of the partitions, the better the results. Beyond these results, we motivate future work to investigate higher values for $i$, which we expect to show even better values, continuing the evident trend in the plot, especially for larger problem instances. However, to ensure comparability between SHA and the selected baselines regarding their runtimes, we conduct the rest of this evaluation using the nodewise approach.


\subsection{Solution Quality}
\begin{figure*}[ht]
    \begin{center}
    \begin{subfigure}[t]{0.49\textwidth}
         \centering
         \includegraphics[width=\columnwidth]{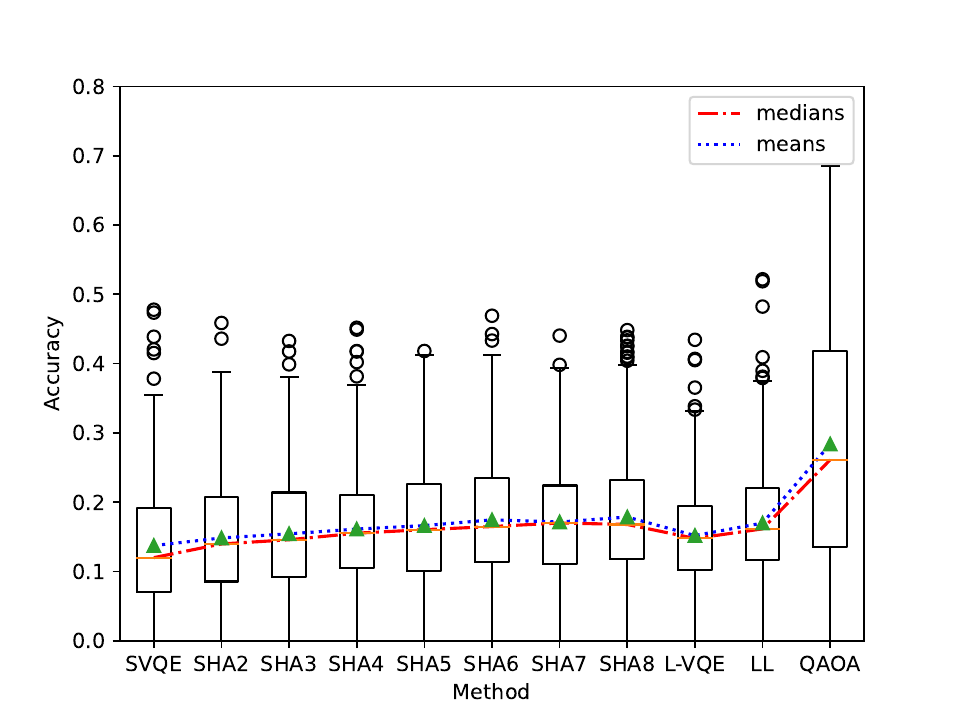}
         \caption{Ratio of correct solutions.}
         \label{fig:accuracy}
     \end{subfigure}
     \begin{subfigure}[t]{0.49\textwidth}
         \centering
         \includegraphics[width=\columnwidth]{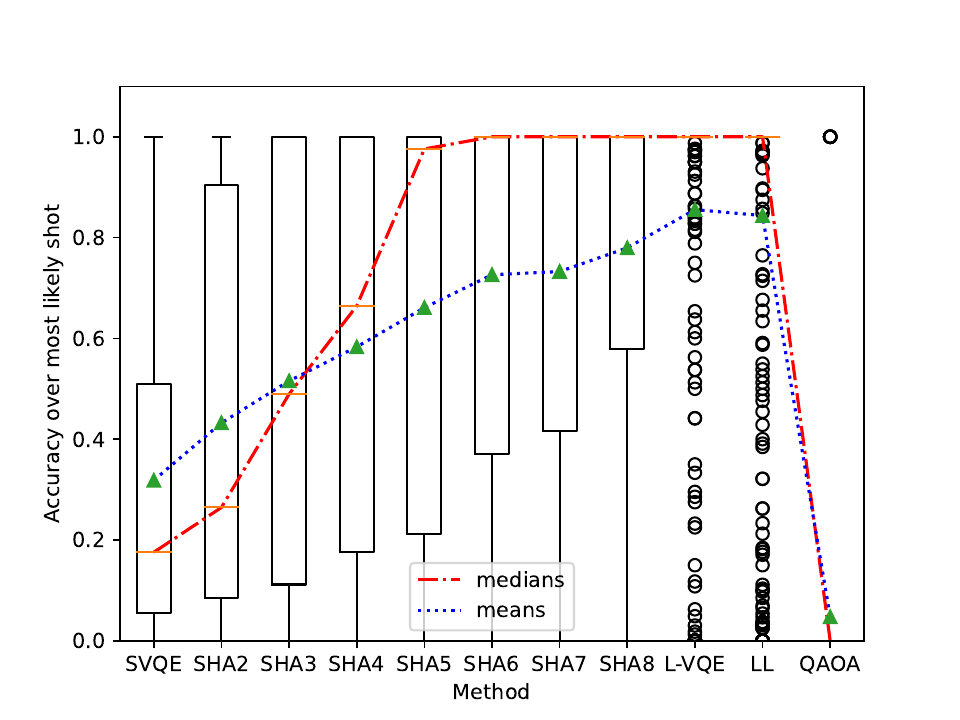}
         \caption{Ratio of correct solutions of the most likely shot, averaged over the last 2\% of optimization steps.}
         \label{fig:accuracy-most-likely-averaged}
     \end{subfigure}
    \caption[LoF entry]{Solution quality averaged over all graphs and circuit architectures for all considered approaches.}
    \label{fig:solution-quality}
    \end{center}
\end{figure*}

To evaluate the solution quality, we examine two properties: The overall accuracy, and the accuracy of the most likely solution. While most literature typically restricts its evaluation to the overall accuracy, additionally investigating the most likely solution has the advantage of revealing the focal point of the identified solution set. Interestingly, our experiments show that these two properties do not automatically coincide, i.e., a comparatively good accuracy for one of both does not indicate a comparatively good accuracy for the other, as shown in figure~\ref{fig:solution-quality} and discussed in the following. Note, that the choice to average over the last couple of optimization steps in figure~\ref{fig:accuracy-most-likely-averaged} allows to see how stable the most likely shot at solving the problem does yield a correct result.

Starting with a closer examination of the results plotted in figure~\ref{fig:accuracy}, we observe, that all methods exceed the standard VQE baseline (SVQE) significantly, with SHA8 showing an enormous 29.99\% improvement. Furthermore, SHA consistently outperforms Layer-VQE (L-VQE) and Layerwise Learning (LL) in terms of raw accuracy, when selecting enough partitions. More concretely, SHA8 displays a 17.58\% better mean accuracy than LL and performs 5.12\% better than L-VQE in the mean, which strongly shows the effectiveness of our proposed approach. However, while it does not influence the results and meaningfulness of this paper, the QAOA baseline still performs significantly better than all VQE-based approaches.

Interestingly, QAOA consistently performs the worst when focusing on the most likely shot (see figure~\ref{fig:accuracy-most-likely-averaged}), while a clear trend of increasing accuracy is visible for the SHA approaches. These results indicate, that while the resulting state vector of the QAOA has many superposition states resembling correct solutions, they are more widespread among the incorrect ones, with a significantly less pronounced peak than the VQE based approaches. While L-VQE and LL strongly focus on such a peak, SHA behaves more volatile.

Concluding these results, we assess that the decision which approach performs the best depends on the practical use case and the available hardware capabilities. When the hardware restrictions allow it, the QAOA can be used and then clearly performs the best in terms of overall accuracy in our low layer depth case study. Otherwise, training a VQE with SHA yields the best overall accuracy. If it is of essence that the distribution of identified solutions has a pronounced, stable, peak at a correct solution, the layerwise learning based VQEs perform almost perfectly, while the QAOA shows significantly worse performance. Future work will have to investigate how this scales for deeper PQCs.

\subsection{Combining SHA with other Quantum Learning Methods}
\begin{figure*}[ht]
    \begin{center}
    \begin{subfigure}[t]{0.49\textwidth}
         \centering
         \includegraphics[width=\columnwidth]{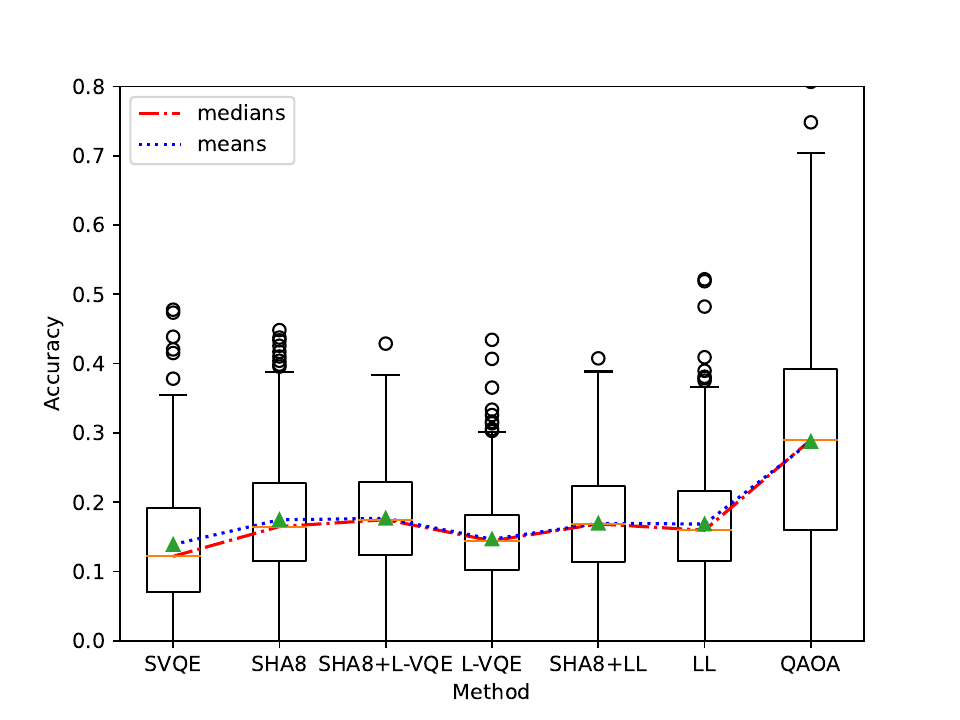}
    \caption{Ratio of correct solutions.}
    \label{fig:accuracy-comb}
     \end{subfigure}
     \begin{subfigure}[t]{0.49\textwidth}
         \centering
         \includegraphics[width=\columnwidth]{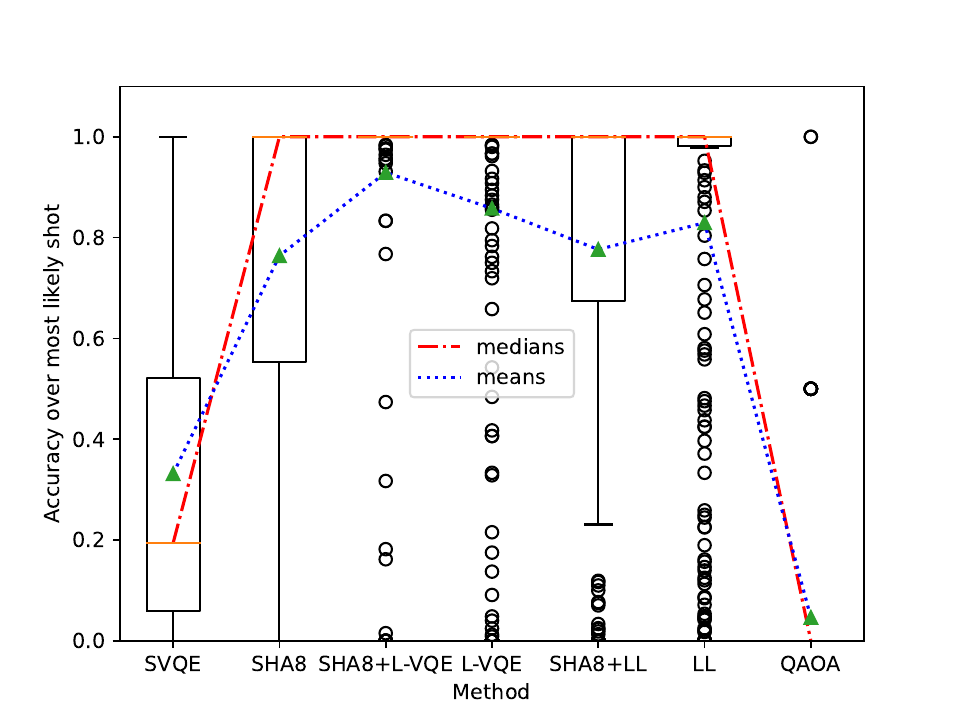}
    \caption{Ratio of correct solutions of the most likely shot, averaged over the last 3\% of optimization steps.}
    \label{fig:accuracy-most-likely-averaged-comb}
     \end{subfigure}
    \caption[LoF entry]{Solution quality averaged over all graphs and circuit architectures when combining different approaches.}
    \label{fig:solution-quality-comb}
    \end{center}
\end{figure*}

As the learning technique employed in SHA is restricted to changes in the cost function, it can be combined with the layerwise learning approaches, which only alter the PQC or the set of trainable parameters. In this section, we evaluate the most straightforward approach to combine SHA with LL and L-VQE, i.e., using SHA for training each newly added circuit layer. Analogously to the previous evaluation on solution quality, we examine the overall accuracy as well as the accuracy of the most likely shot, as displayed in figure~\ref{fig:solution-quality-comb}.

In terms of the overall accuracy (aside from the QAOA), the combination of SHA8 and L-VQE achieve the best results with a median accuracy improvement of 35.5\% against the standard VQE (SVQE). This reveals a powerful synergy of combining SHA and L-VQE, as L-VQE performs worse than LL when not coupled with SHA, but does so for their hybrid variants. Overall the SHA+V-VQE hybrid outperforms the previous empirical state of the art VQE approach (i.e., LL), by 5\% in the mean.  Examining the results for the most likely shot, displayed in figure~\ref{fig:accuracy-most-likely-averaged-comb}, we can see that the SHA+L-VQE hybrid also performs the best overall, while improving 8.31\% over the previous best mean result (L-VQE). Interestingly, the SHA+LL hybrid does only perform slightly better than SHA (1.63\%), but is worse than LL in this metric. While the exact reasons for this performance difference remain unclear based on the conducted experiments, this shows that hybrid approaches do not automatically improve the overall performance, but can also even have negative effects on it.

\subsection{Time Complexity}
Aside from solution quality, the training duration is a crucial performance indicator in practice. The less optimization steps are needed, the faster the training process in practice (i.e., when considering a similar amount of parameters). The similarity in the number of parameters is important, because the quantum gradient calculation scales linearly in this entity (see section~\ref{subsec:prametertraining}). As the actual number of concurrently trained parameters is generally smaller in the layerwise approaches, this must be accounted for in an exact runtime analysis. On average, our implementation of LL trains only about half the number of parameters in the full PQC, while L-VQE trains roughly $\nicefrac{2}{3}$. As the circuit depth also changes in layerwise learning approaches (and with that, the time needed to execute the circuits), this also influences the training time complexity. Using our choice of hyperparameters, LL executes $\nicefrac{3}{4}$ of the full PQC on average, while L-VQE only executes $\nicefrac{2}{3}$.

\begin{figure}[ht]
    \centering
    \includegraphics[width=\columnwidth]{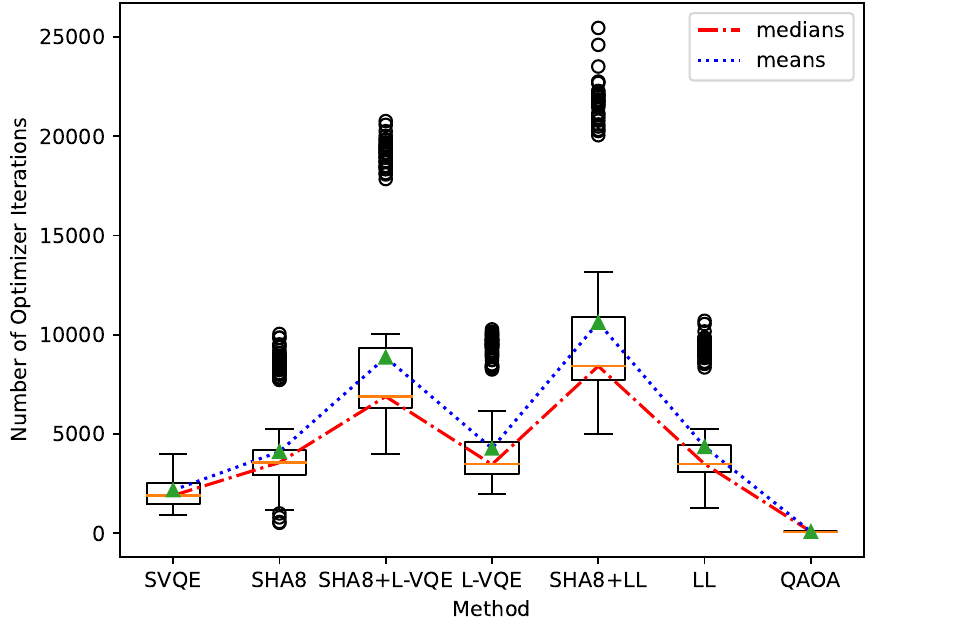}
    \caption{Number of optimization iterations.}
    \label{fig:speed-w/pre}
\end{figure}

However, due to the small absolute number of parameters and layers in our setting, we softly neglect these complicating factors and inspect the raw difference in the number of optimization iterations as displayed in figure~\ref{fig:speed-w/pre}. When comparing SHA to the standard VQE, we observe a close to doubled number of optimization iterations, which indicates, that the significant improvement in solution quality comes at the cost of longer training time. However, real world applications exist, in which the better achievable solution quality outweighs the increase in runtime, as, e.g., when benchmarking for the best possible performance of a given VQE while aiming to show early quantum advantage on a given QPU. Following the considerations above, L-VQE and LL will have fast wallclock times than SHA, even though they are on par in the number of optimization iterations. For the hybrid approaches, a significant increase in runtime can be observed, limiting their practical use cases in spite of the qualitative improvements. When comparing the SHA+LL hybrid to the SHA+L-VQE hybrid, it becomes apparent that the approach with the better solution quality (SHA+L-VQE) also trains faster, which indicates, that the parameter landscape is easier to navigate through. Finally, it has to be acknowledged, that a comparison of the VQE-based approaches to the QAOA baseline is especially hard, since QAOA's PQC only has six parameters, accelerating the optimization routine immensely. Nevertheless, this even underscores, the tremendously faster trainability of the QAOA, at least in the regime of short PQCs.

\section{\uppercase{Conclusion}}
\label{sec:conclusion}
The goal of this contribution was the development of a novel quantum learning method, that facilitates the parameter training for VQEs. Stemming on the fact that increased locality in the cost function decreases the risk of encountering vanishing gradients, we proposed the sequential Hamiltonian assembly (SHA) technique, that iteratively approximates the possibly global cost function by assembling it from its local components. In our experiments, we showed that our approach improves the mean solution quality of the standard VQE approach by 29.99\% and outperforms the empirical state of the art Layer-VQE by 5.12\%. This demonstrates a prove of principle and motivates further research on parameter training approaches, that iteratively approximate the original cost function to guide the learning process. To extend this case study analysis of our approach, other problems beyond graph coloring should be tested. An interesting candidate might be Max-Cut, as it also offers a graph structure, allowing to examining the performance the proposed nodewise assembly strategy. Beyond this, exploring new strategies for graph and non-graph problems invites for future work. As SHA slightly increases the wallclock training time in its evaluated form, a more extensive hyperparameter search should be conducted to explore better quality/time tradeoffs. Having seen that combinations of existing layerwise learning approaches with SHA can increase the solution quality even further, gathering more information on their interplay might open up new approaches of quantum learning methods, that are especially targeted towards attacking multiple issues of vanishing gradients concurrently, as, in our case locality and expressiveness. In conclusion, our contribution uncovers a new, locality based, approach towards efficiently learning parameters in parameterized quantum circuits.

\section*{\uppercase{Acknowledgements}} 
This paper was partially funded by the German Federal Ministry for Economic Affairs and Climate Action through the funding program "Quantum Computing -- Applications for the industry" based on the allowance "Development of digital technologies" (contract number: 01MQ22008A).

\bibliographystyle{apalike}
{\small
\bibliography{main}}

\begin{thebibliography}{}

\bibitem[Aharonov et~al., 2004]{10.1109/FOCS.2004.8}
Aharonov, D., van Dam, W., Kempe, J., Landau, Z., Lloyd, S., and Regev, O. (2004).
\newblock Adiabatic quantum computation is equivalent to standard quantum computation.
\newblock In {\em Proceedings of the 45th Annual IEEE Symposium on Foundations of Computer Science}, FOCS '04, page 42–51, USA. IEEE Computer Society.

\bibitem[Albash and Lidar, 2018]{PhysRevX.8.031016}
Albash, T. and Lidar, D.~A. (2018).
\newblock Demonstration of a scaling advantage for a quantum annealer over simulated annealing.
\newblock {\em Phys. Rev. X}, 8:031016.

\bibitem[Bansal, 2014]{bansal:LIPIcs:2014:4827}
Bansal, N. (2014).
\newblock {New Developments in Iterated Rounding (Invited Talk)}.
\newblock In Raman, V. and Suresh, S.~P., editors, {\em 34th International Conference on Foundation of Software Technology and Theoretical Computer Science (FSTTCS 2014)}, volume~29 of {\em Leibniz International Proceedings in Informatics (LIPIcs)}, pages 1--10, Dagstuhl, Germany. Schloss Dagstuhl--Leibniz-Zentrum fuer Informatik.

\bibitem[Bengio et~al., 2006]{NIPS2006_5da713a6}
Bengio, Y., Lamblin, P., Popovici, D., and Larochelle, H. (2006).
\newblock Greedy layer-wise training of deep networks.
\newblock In Sch\"{o}lkopf, B., Platt, J., and Hoffman, T., editors, {\em Advances in Neural Information Processing Systems}, volume~19. MIT Press.

\bibitem[Blekos et~al., 2023]{blekos2023review}
Blekos, K., Brand, D., Ceschini, A., Chou, C.-H., Li, R.-H., Pandya, K., and Summer, A. (2023).
\newblock A review on quantum approximate optimization algorithm and its variants.

\bibitem[Born and Fock, 1928]{Born1928}
Born, M. and Fock, V. (1928).
\newblock {Beweis des Adiabatensatzes}.
\newblock {\em Zeitschrift f{\"{u}}r Phys.}, 51(3):165--180.

\bibitem[Bradley, 2010]{bradley2010learning}
Bradley, D.~M. (2010).
\newblock {\em Learning in modular systems}.
\newblock Carnegie Mellon University.

\bibitem[Campos et~al., 2021]{PhysRevA.103.032607}
Campos, E., Nasrallah, A., and Biamonte, J. (2021).
\newblock Abrupt transitions in variational quantum circuit training.
\newblock {\em Phys. Rev. A}, 103:032607.

\bibitem[Cerezo et~al., 2021a]{Cerezo2021}
Cerezo, M., Arrasmith, A., Babbush, R., Benjamin, S.~C., Endo, S., Fujii, K., McClean, J.~R., Mitarai, K., Yuan, X., Cincio, L., and Coles, P.~J. (2021a).
\newblock {Variational quantum algorithms}.
\newblock {\em Nat. Rev. Phys.}, 3(9):625--644.

\bibitem[Cerezo et~al., 2021b]{cerezo2021cost}
Cerezo, M., Sone, A., Volkoff, T., Cincio, L., and Coles, P.~J. (2021b).
\newblock Cost function dependent barren plateaus in shallow parametrized quantum circuits.
\newblock {\em Nature communications}, 12(1):1791.

\bibitem[Deutsch and Jozsa, 1992]{deutsch1992rapid}
Deutsch, D. and Jozsa, R. (1992).
\newblock Rapid solution of problems by quantum computation.
\newblock {\em Proceedings of the Royal Society of London. Series A: Mathematical and Physical Sciences}, 439(1907):553--558.

\bibitem[Du et~al., 2022]{Du2022}
Du, Y., Huang, T., You, S., Hsieh, M.-H., and Tao, D. (2022).
\newblock {Quantum circuit architecture search for variational quantum algorithms}.
\newblock {\em npj Quantum Inf.}, 8(1):62.

\bibitem[Farhi et~al., 2014]{farhi2014quantum}
Farhi, E., Goldstone, J., and Gutmann, S. (2014).
\newblock A quantum approximate optimization algorithm.

\bibitem[Fontana et~al., 2023]{fontana2023adjoint}
Fontana, E., Herman, D., Chakrabarti, S., Kumar, N., Yalovetzky, R., Heredge, J., Sureshbabu, S.~H., and Pistoia, M. (2023).
\newblock The adjoint is all you need: Characterizing barren plateaus in quantum ans\"atze.

\bibitem[Gilbert, 1959]{10.1214/aoms/1177706098}
Gilbert, E.~N. (1959).
\newblock {Random Graphs}.
\newblock {\em The Annals of Mathematical Statistics}, 30(4):1141 -- 1144.

\bibitem[Glorot and Bengio, 2010]{pmlr-v9-glorot10a}
Glorot, X. and Bengio, Y. (2010).
\newblock Understanding the difficulty of training deep feedforward neural networks.
\newblock In Teh, Y.~W. and Titterington, M., editors, {\em Proceedings of the Thirteenth International Conference on Artificial Intelligence and Statistics}, volume~9 of {\em Proceedings of Machine Learning Research}, pages 249--256, Chia Laguna Resort, Sardinia, Italy. PMLR.

\bibitem[Gomez et~al., 1994]{COBYLA_1}
Gomez, S., Hennart, J.-P., and Powell, M. J.~D., editors (1994).
\newblock {\em A Direct Search Optimization Method That Models the Objective and Constraint Functions by Linear Interpolation}, pages 51--67.
\newblock Springer Netherlands, Dordrecht.

\bibitem[Grover, 1996]{10.1145/237814.237866}
Grover, L.~K. (1996).
\newblock A fast quantum mechanical algorithm for database search.
\newblock In {\em Proceedings of the Twenty-Eighth Annual ACM Symposium on Theory of Computing}, STOC '96, page 212–219, New York, NY, USA. Association for Computing Machinery.

\bibitem[Hagberg et~al., 2008]{hagberg2008exploring}
Hagberg, A., Swart, P., and S~Chult, D. (2008).
\newblock Exploring network structure, dynamics, and function using networkx.
\newblock Technical report, Los Alamos National Lab.(LANL), Los Alamos, NM (United States).

\bibitem[Holmes et~al., 2022]{PRXQuantum.3.010313}
Holmes, Z., Sharma, K., Cerezo, M., and Coles, P.~J. (2022).
\newblock Connecting ansatz expressibility to gradient magnitudes and barren plateaus.
\newblock {\em PRX Quantum}, 3:010313.

\bibitem[Huang et~al., 2020]{9345015}
Huang, Y., Lei, H., and Li, X. (2020).
\newblock An empirical study of optimizers for quantum machine learning.
\newblock In {\em 2020 IEEE 6th International Conference on Computer and Communications (ICCC)}, pages 1560--1566.

\bibitem[Joshi et~al., 2021]{Joshi_2021}
Joshi, N., Katyayan, P., and Ahmed, S.~A. (2021).
\newblock Evaluating the performance of some local optimizers for variational quantum classifiers.
\newblock {\em Journal of Physics: Conference Series}, 1817(1):012015.

\bibitem[Kashif and Al-Kuwari, 2023]{kashif2023impact}
Kashif, M. and Al-Kuwari, S. (2023).
\newblock The impact of cost function globality and locality in hybrid quantum neural networks on nisq devices.
\newblock {\em Machine Learning: Science and Technology}, 4(1):015004.

\bibitem[Knill et~al., 2007]{knill2007optimal}
Knill, E., Ortiz, G., and Somma, R.~D. (2007).
\newblock Optimal quantum measurements of expectation values of observables.
\newblock {\em Physical Review A}, 75(1):012328.

\bibitem[Lanczos, 2012]{lanczos2012variational}
Lanczos, C. (2012).
\newblock {\em The Variational Principles of Mechanics}.
\newblock Dover Books on Physics. Dover Publications.

\bibitem[Liu et~al., 2022]{9669165}
Liu, X., Angone, A., Shaydulin, R., Safro, I., Alexeev, Y., and Cincio, L. (2022).
\newblock Layer vqe: A variational approach for combinatorial optimization on noisy quantum computers.
\newblock {\em IEEE Transactions on Quantum Engineering}, 3:1--20.

\bibitem[Lucas, 2014]{10.3389/fphy.2014.00005}
Lucas, A. (2014).
\newblock Ising formulations of many np problems.
\newblock {\em Frontiers in Physics}, 2.

\bibitem[McClean et~al., 2018]{McClean2018}
McClean, J.~R., Boixo, S., Smelyanskiy, V.~N., Babbush, R., and Neven, H. (2018).
\newblock {Barren plateaus in quantum neural network training landscapes}.
\newblock {\em Nat. Commun.}, 9(1):4812.

\bibitem[Mitarai et~al., 2018]{PhysRevA.98.032309}
Mitarai, K., Negoro, M., Kitagawa, M., and Fujii, K. (2018).
\newblock Quantum circuit learning.
\newblock {\em Phys. Rev. A}, 98:032309.

\bibitem[Nielsen and Chuang, 2010]{nielsen_chuang_2010}
Nielsen, M.~A. and Chuang, I.~L. (2010).
\newblock {\em Quantum Computation and Quantum Information: 10th Anniversary Edition}.
\newblock Cambridge University Press.

\bibitem[Peruzzo et~al., 2014]{Peruzzo2014}
Peruzzo, A., McClean, J., Shadbolt, P., Yung, M.-H., Zhou, X.-Q., Love, P.~J., Aspuru-Guzik, A., and O'Brien, J.~L. (2014).
\newblock {A variational eigenvalue solver on a photonic quantum processor}.
\newblock {\em Nat. Commun.}, 5(1):4213.

\bibitem[Pirnay et~al., 2023]{pirnay2023inprinciple}
Pirnay, N., Ulitzsch, V., Wilde, F., Eisert, J., and Seifert, J.-P. (2023).
\newblock An in-principle super-polynomial quantum advantage for approximating combinatorial optimization problems.

\bibitem[Ragone et~al., 2023]{ragone2023unified}
Ragone, M., Bakalov, B.~N., Sauvage, F., Kemper, A.~F., Marrero, C.~O., Larocca, M., and Cerezo, M. (2023).
\newblock A unified theory of barren plateaus for deep parametrized quantum circuits.

\bibitem[Sack et~al., 2022]{PRXQuantum.3.020365}
Sack, S.~H., Medina, R.~A., Michailidis, A.~A., Kueng, R., and Serbyn, M. (2022).
\newblock Avoiding barren plateaus using classical shadows.
\newblock {\em PRX Quantum}, 3:020365.

\bibitem[Schuld et~al., 2019]{PhysRevA.99.032331}
Schuld, M., Bergholm, V., Gogolin, C., Izaac, J., and Killoran, N. (2019).
\newblock Evaluating analytic gradients on quantum hardware.
\newblock {\em Phys. Rev. A}, 99:032331.

\bibitem[Schuld et~al., 2021]{PhysRevA.103.032430}
Schuld, M., Sweke, R., and Meyer, J.~J. (2021).
\newblock Effect of data encoding on the expressive power of variational quantum-machine-learning models.
\newblock {\em Phys. Rev. A}, 103:032430.

\bibitem[Sim et~al., 2019a]{sim2019expressibility}
Sim, S., Johnson, P.~D., and Aspuru-Guzik, A. (2019a).
\newblock Expressibility and entangling capability of parameterized quantum circuits for hybrid quantum-classical algorithms.
\newblock {\em Advanced Quantum Technologies}, 2(12):1900070.

\bibitem[Sim et~al., 2019b]{https://doi.org/10.1002/qute.201900070}
Sim, S., Johnson, P.~D., and Aspuru-Guzik, A. (2019b).
\newblock Expressibility and entangling capability of parameterized quantum circuits for hybrid quantum-classical algorithms.
\newblock {\em Advanced Quantum Technologies}, 2(12):1900070.

\bibitem[Skolik et~al., 2021]{Skolik2021}
Skolik, A., McClean, J.~R., Mohseni, M., van~der Smagt, P., and Leib, M. (2021).
\newblock {Layerwise learning for quantum neural networks}.
\newblock {\em Quantum Mach. Intell.}, 3(1):5.

\bibitem[Stilck~Fran{\c{c}}a and Garcia-Patron, 2021]{stilck2021limitations}
Stilck~Fran{\c{c}}a, D. and Garcia-Patron, R. (2021).
\newblock Limitations of optimization algorithms on noisy quantum devices.
\newblock {\em Nature Physics}, 17(11):1221--1227.

\bibitem[Tabi et~al., 2020]{9259934}
Tabi, Z., El-Safty, K.~H., Kallus, Z., Hága, P., Kozsik, T., Glos, A., and Zimborás, Z. (2020).
\newblock Quantum optimization for the graph coloring problem with space-efficient embedding.
\newblock In {\em 2020 IEEE International Conference on Quantum Computing and Engineering (QCE)}, pages 56--62.

\bibitem[Uvarov and Biamonte, 2021]{uvarov2021barren}
Uvarov, A. and Biamonte, J.~D. (2021).
\newblock On barren plateaus and cost function locality in variational quantum algorithms.
\newblock {\em Journal of Physics A: Mathematical and Theoretical}, 54(24):245301.

\bibitem[Wang et~al., 2021]{wang2021noise}
Wang, S., Fontana, E., Cerezo, M., Sharma, K., Sone, A., Cincio, L., and Coles, P.~J. (2021).
\newblock Noise-induced barren plateaus in variational quantum algorithms.
\newblock {\em Nature communications}, 12(1):6961.

\bibitem[Zdeborov\'a and Krz\k{a}ka\l{}a, 2007]{PhysRevE.76.031131}
Zdeborov\'a, L. and Krz\k{a}ka\l{}a, F. (2007).
\newblock Phase transitions in the coloring of random graphs.
\newblock {\em Phys. Rev. E}, 76:031131.

\bibitem[Zhang et~al., 2022]{NEURIPS2022_7611a3cb}
Zhang, K., Liu, L., Hsieh, M.-H., and Tao, D. (2022).
\newblock Escaping from the barren plateau via gaussian initializations in deep variational quantum circuits.
\newblock In Koyejo, S., Mohamed, S., Agarwal, A., Belgrave, D., Cho, K., and Oh, A., editors, {\em Advances in Neural Information Processing Systems}, volume~35, pages 18612--18627. Curran Associates, Inc.

\end{thebibliography}


\end{document}